\documentclass[aps,footinbib,twocolumn,showpacs,amsmath,amssymb,superscriptaddress]{revtex4}
\usepackage{graphicx,amsmath}
\usepackage{epstopdf}
\usepackage{amssymb}
\usepackage{grffile}
\usepackage[usenames]{color}
\usepackage{indentfirst}
\usepackage{float}
\usepackage{color}
\usepackage{comment}
\usepackage{mathrsfs}
\usepackage{dcolumn}
\usepackage{bm}
\usepackage[colorlinks=true,bookmarks=false,citecolor=blue,linkcolor=red,urlcolor=blue]{hyperref}

\begin{document} 
\title{The structure of heavily doped impurity band in crystalline host} 
\author{Hongwei Chen}
\affiliation{Department of Physics, Northeastern University, Boston, Massachusetts 02115}
\affiliation{Stanford Institute for Materials and Energy Sciences, Stanford University, Stanford, CA 94305}
\affiliation{Linac Coherent Light Source, SLAC National Accelerator Laboratory, Menlo Park, CA 94720}

\author{Zi-Xiang Hu}
\email{zxhu@cqu.edu.cn}
\affiliation{Department of Physics and Chongqing Key Laboratory for Strongly Coupled Physics, Chongqing University, Chongqing 401331, People's Republic of China}

\begin{abstract}
We study the properties of the impurity band in heavily-doped non-magnetic semiconductors using the Jacobi-Davidson algorithm and the supervised deep learning method. The disorder averaged inverse participation ratio (IPR) and thouless number calculation show us the rich structure inside the impurity band. A Convolutional Neural Network(CNN) model, which is trained to distinguish the extended/localized phase of the Anderson model with high accuracy, shows us the results in good agreement with the conventional approach. Together, we find that there are three mobility edges in the impurity band for a specific on-site impurity potential, which means the presence of the extended states while filling the impurity band.
\end{abstract}
\pacs{71.23.-k, 71.55.-i, 02.60.Cb}
\date{\today}
\maketitle 
 
\section{Introduction}
The effect of disorder has been extensively studied since Anderson's seminal paper\cite{anderson1958absence}. Diluted magnetic semiconductors (DMS) doped with a small concentration of charged impurities constitute an interesting magnetic system that has a number of novel features for study by numerical simulation\cite{Avrous1991semimagneticSA}. Much of the research has been focused on II-VI (such as CdTe or ZnSe) and III-V (such as GaAs) compound semiconductors doped with a low concentration ($x\sim 1-8\%$) of Manganese (Mn) impurities. Of particular interest in this field is Ga$_{1-x}$Mn${_x}$As which has been shown to exhibit ferromagnetic behavior above 100K\cite{ohno1998making}. In these samples, the Manganese is substitutions with the Gallium and acts as an acceptor (donating one hole to the crystal), so that the material is p-type. The holes bind to the impurities with an energy of around 130 meV around $x\sim 10\%$\cite{beschoten1999magnetic}. Since $x \ll 1 $, the overlap between different impurity states can be ignored, thus the interaction between the charge carriers can be neglected. The system can be simply described by a noninteracting tight-binding model. When the system contains only one impurity, and the binding energy is large enough, an impurity state appears below the conductance band (we assume the impurity potential is attractive). It is locally distributed in space near the impurity potential within a localization length $\zeta$. As increasing the concentration $x$, the overlap between different impurity states extends the single impurity energy to an impurity band in the density of state (DOS) and eventually merges with the conductance band. Simultaneously, the states in the impurity band are expected to become more and more extended and ultimately regain their bandlike character~\cite{Mona}. However, the details inside the impurity band are rarely studied.

One reason for lacking such a study is the computation difficulty even in the non-interacting case. Generally, the percentage of the state in the impurity band in the total number of states is about $10\% $ at the concentration we are interested in. Taking a 3-dimensional Anderson model with lattice size $30\times 30\times 30$ as an example, the number of states which we need to know in the impurity band is about 3000. The exact diagonalization\cite{weisse2008exact} for such a system is very difficult due to the large dimension. On the other hand, we have to do a large number of sample averages. The sparse matrix diagonalization, such as the Lanczos method\cite{lanczos_paper}, can be adapted to obtain a few lowest-lying states or a few states nearby special energy (the simplest way is diagonalizing $(H - \epsilon I)^2$ by using the original Lanczos diagonalization method).

Machine learning methods have recently emerged as a valuable tool to study the quantum many-body physics problems\cite{Carleo2017, carrasquilla2017machine, ch2017machine, van2017learning, PhysRevLett.120.257204, PhysRevE.96.022140, rodriguez2019identifying, lidiak2020unsupervised,  hsu2018machine, hendry2021chebyshev, choo2019two, pfau2020ab, sharir2020deep, hendry2022thermal, chen2022systematic}. Its ability to process high dimensional data and recognize complex patterns have been utilized to determine phase diagrams and phase transitions\cite{PhysRevB.94.195105, ohtsuki2017deep, tanaka2017detection, mano2017phase, broecker2017machine,  schindler2017probing, PhysRevB.99.075418, PhysRevB.99.121104, kotthoff2021distinguishing, zhang2019interpretable, PhysRevE.99.032142, K_ming_2021}. In particular, Convolutional Neural Network(CNN)\cite{alexnet} model, which initially is designed for image recognition, was widely used to study different kinds of phase transition problems including the Bose-Hubbard model\cite{bohrdt2021analyzing}, spin 1/2 Heisenberg model\cite{theveniaut2019neural}, quantum transverse-field Ising model\cite{zhang2019interpretable} and etc. The power of using machine learning to recognize quantum states lies in their ability to finish tasks without the knowledge of physics background or the Hamiltonian of the system. Even if the neural network is trained in a small energy region of the system, it can be used to obtain the whole phase diagram\cite{ohtsuki2017deep, mano2017phase}. Also, it can discriminate quantum states with high accuracy even if they are trained from a totally different Hamiltonian. This special feature of machine learning inspires us to try to identify the delocalized states in the “impurity band”.

In this paper, we develop a method to obtain the correct density of states (DOS) and other localization properties, such as inverse participation ratio (IPR)\cite{brndiar2006universality} and thouless number\cite{edwards1972numerical}, by using Jacobi-Davidson sparse matrix diagonalization\cite{Bollhfer2007JADAMILUAS} with an importance sampling statistics method. Meanwhile, we train a 3-dimensional CNN model using the data generated from the Anderson model, and then the trained model is used to identify the existence of extended states in the impurity band. This manuscript is organized as follows: In sec.\ref{sec:model} we describe the tight-binding model on the cubic lattice and numerical methods; Sec. \ref{sec:doping} demonstrates the effect of heavy doping studied by studying the IPR and Thouless number; Sec.\ref{sec:ml} demonstrates the implementation of the deep learning approach and the results from the trained neural network model;  finally, we close with a conclusion.

\section{Model and Methods} \label{sec:model}

We consider a tight-binding model on a D-dimensional hypercubic lattice with the nearest neighbor hopping t, and on-site energies $\epsilon_i$:
\begin{equation}
    H = -t \sum_{\langle i, j \rangle}(\hat{c}^{\dagger}_j \hat{c}_j + h.c.) + \sum_i \epsilon_i \hat{c}^{\dagger}_i \hat{c}_i
\label{eq:hamiltonian}
\end{equation}
The hopping term simulates the iterative electrons and the on-site energy has a bimodal distribution $\epsilon_i = -W$ with probability $x$, and $\epsilon_i = 0$ with probability ($1-x$). This model a host lattice with a single relevant band, with a fraction $x$ of substitutional impurities. For one-dimensional (d = 1) free electrons, the energy-momentum dispersion relation is $E(k) = 2t\cos(k)$, it is easy to get the DOS with the formula

\begin{equation}
    \rho(E) = (\frac{1}{2\pi})^d \int \frac{dS}{\nabla_k E}.
\end{equation}
The result for 1D is:
\begin{equation}
    \rho_{1d}(E) =   \frac{1}{\sqrt{4t^2 - E^2}}.
\end{equation}
\begin{figure}
 \includegraphics[width=8cm]{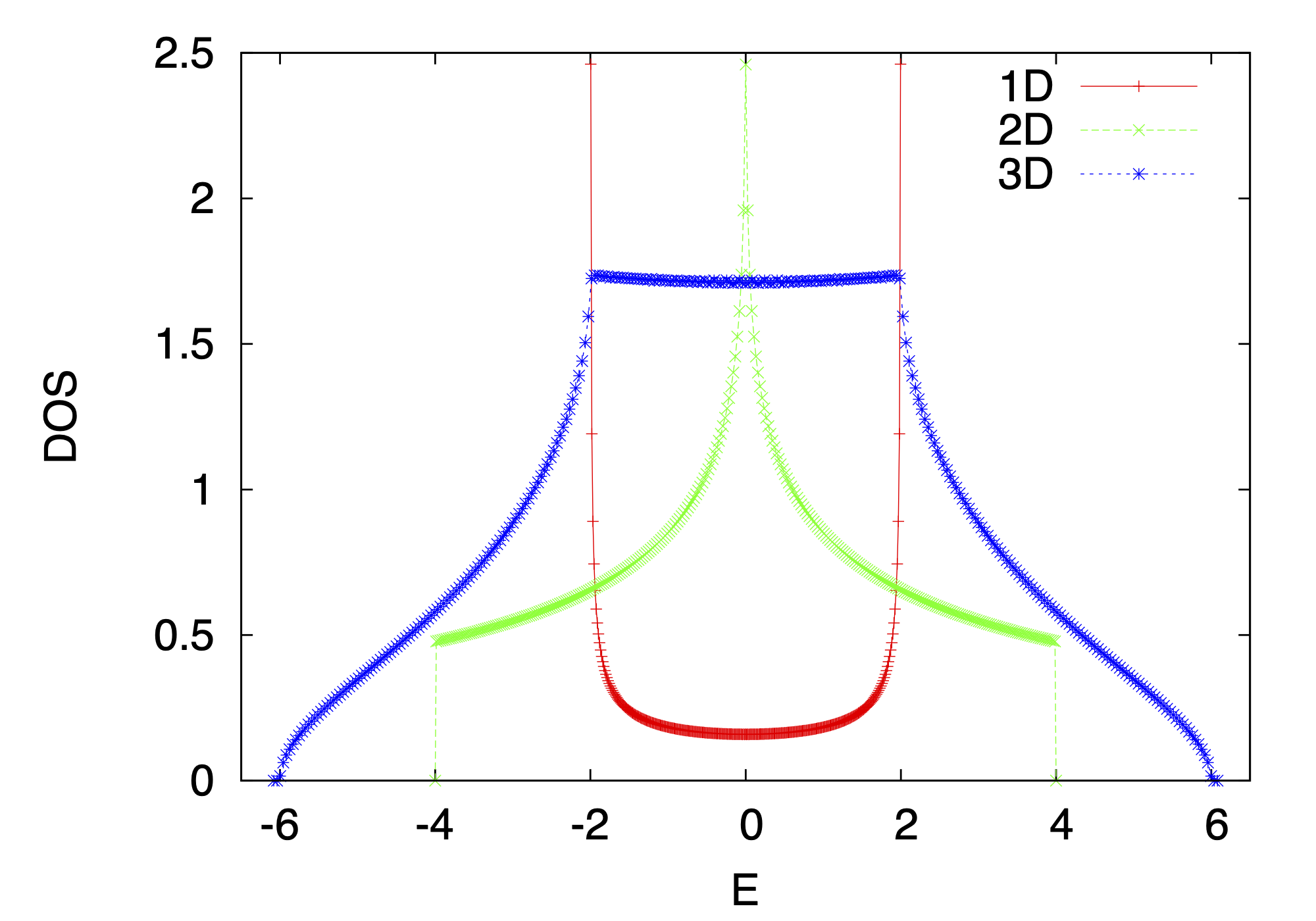}
\caption{\label{dosfree}The density of states for free electrons in the tight-binding model in one, two, and three dimensions. Here $t$ has been set to be unit.}
\end{figure}
There is no analytic solution for higher dimensional systems, however, an approximation that is accurate to roughly $2\%$ was given by Andres et al~\cite{andres1981low}. Instead, the DOS can be calculated numerically by exact diagonalization as shown in Fig. \ref{dosfree} where $t$ has been set to the unit. After introducing the impurities, all states become localized in 1D and 2D based on the scaling theory of localization~\cite{Abrahams}. Part of the states become to be localized and develop into an impurity band at the edge of the conducting band. To determine the localized/extended state, namely the location of the mobility gap, we calculate the inverse participation ratio (IPR)\cite{brndiar2006universality}
\begin{equation}
    \text{IPR} = \frac{\sum_i |\psi_i^4|}{(\sum_i |\psi_i^2|)^2}
\end{equation}
for each state, where the $\psi_i$ is the weight of an eigen wave function on the $i$'th site. Heuristically, if we compare two trivial states with wave functions for a $N$-site system:
\begin{equation}
    \Psi_{extended} = \sum_i(\psi_i = 1/\sqrt{N}) = \sqrt{N},
\end{equation}
and
\begin{equation}
    \Psi_{localized}(j) = \sum_i(\psi_i = \delta_{ij}) = 1
\end{equation}
where $\Psi_{extended}$ is an extended state which has equal weight on each site and $\Psi_{localized}(j)$ is a localized state which only has weight on the $j$'th site. It is easy to see that the IPR of $\Psi_{extended}$ decreased with the order of $\frac{1}{N}$ and a constant for $\Psi_{localized}(j)$. On the other hand, the Thouless number\cite{edwards1972numerical} is defined as:
\begin{equation}
    g(E) = \frac{\langle |\Delta E| \rangle}{\langle \delta E \rangle},
\end{equation}
where $\delta E$ is the energy difference while the boundary condition changes from periodic boundary condition (PBC) to anti-periodic boundary condition (APBC) and the $|\Delta E|$ is the average energy distance around $E$. Since only the extended states are sensitive to the change of boundary condition, $g(E)$ grows linearly as a function of the system size for the extended state, and conversely, it reduces for the localized state.
In this work, we determine the localization properties by systematically studying the IPR and Thouless number for different system sizes, and the crossover points of the Thouless number give us a hint of the  mobility edge.

\begin{figure}
\includegraphics[width=8cm]{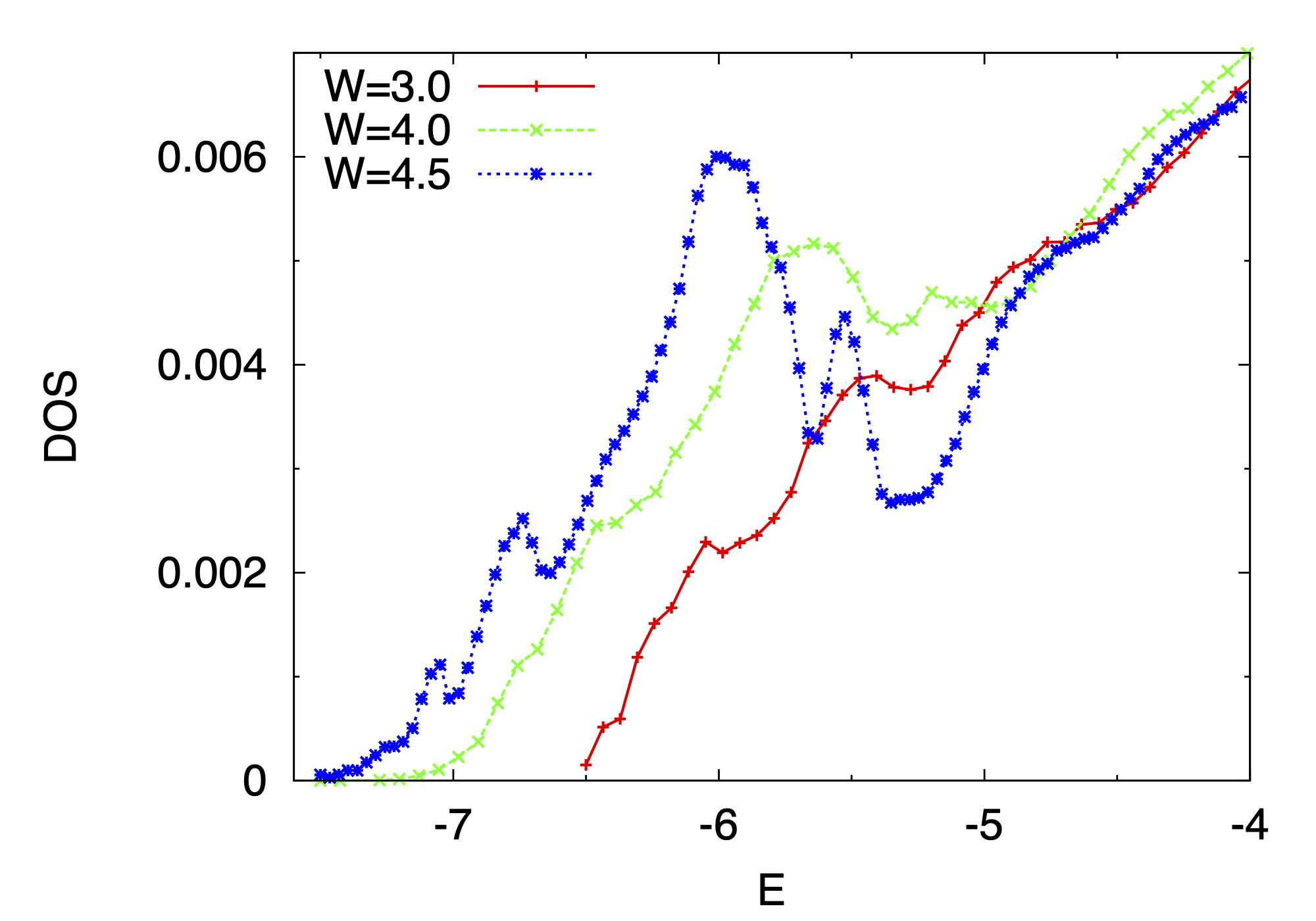}
\caption{\label{DOSEvolution} The evolution of the DOS at the band edge with different doping strengths. The system has size $19\times 20 \times 21$ and can be fully diagonalized.}
\end{figure}

For three dimensional cubic lattice of size $L$, the Hamiltonian matrix has a dimension of $L^3$. General full exact diagonalization methods, such as Lapack library~\cite{lapack99}, can only deal with small system sizes. The computation time of diagonalizing one matrix with size $L^3$ grows dramatically as a function of the system size. As shown in Fig.~\ref{DOSEvolution}, we deal with a system with size $19\times 20 \times 21$ with doping concentration $x = 5\%$, after averaging thousands of samples, we obtained the DOS for different doping energies. It is shown that a peak emerges gradually near the band edge as increasing the doping energy $W$. This peak becomes more prominent around $W \sim 4.5$ at which an obvious depletion is developed at the junction between the impurity band and the conduction band.

 Since only the developed impurity band is the interesting part we are focusing on. The number of states in the impurity band is about the lowest $10\%$ of states in the whole band, thus we do not have to fully diagonalize the Hamiltonian. On the other side, we just need to calculate the DOS, IPR, and Thouless number for these lowest $10\%$ states after averaging thousands of samples. According to our demand, we use the sparse matrix diagonalization with Jacobi-Division (JADA) method~\cite{Bollhfer2007JADAMILUAS} which can search a few (10 to 20) states efficiently near specific points.   For a given sample at fixed doping strength, we randomly distribute the reference points (30-50 points) in the impurity band, taking $W=-4.5$ as an example, the reference points are picked randomly in the region $[-8:-4]$, about 10-20 states can be obtained by JADA around each reference point. The reference points could also be picked by importance sampling based on the DOS for a small system from the full diagonalization. We collect all these energies for each reference point in one sample. After thousands of sample averaging, we obtain the same DOS as that from the full diagonalization for a small system. It is obvious that the JADA method can easily go beyond the limit of the full exact diagonalization. At least on the same price of the computation time, we can nearly double the system size compared to the Lapack method. In this work, we calculate the properties for system sizes up to $40^3$ sites by using the JADA method.
 
\section{The effect of heavily doping} \label{sec:doping}
 \begin{figure}
  \includegraphics[width=8cm]{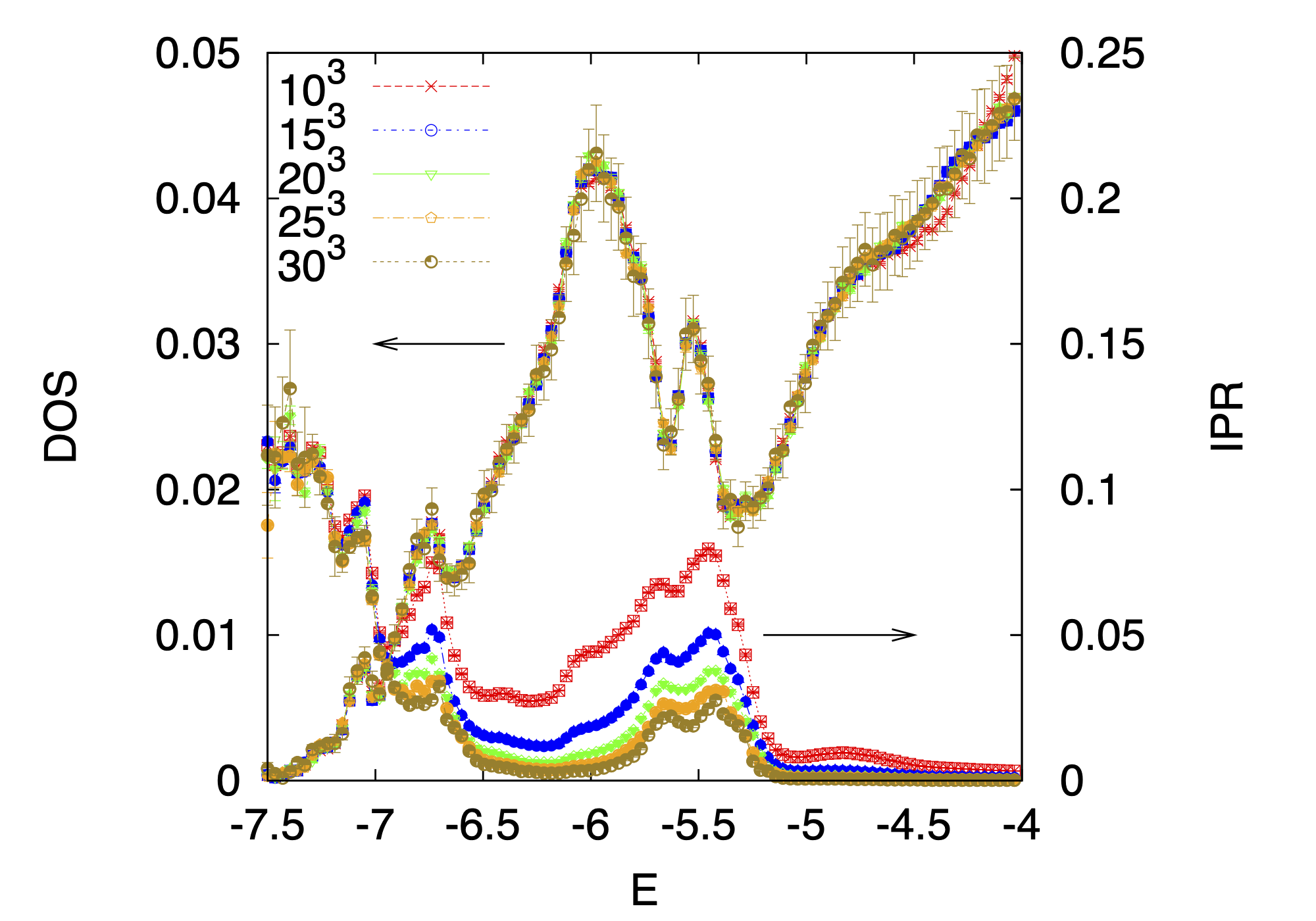}
 \caption{\label{dosipr}The DOS and IPR for $5\%$ doped system with $W=-4.5$. The results are obtained from exact diagonalization. The number of configurations ranges from 1000 for system $14\times15\times16$ to 50 for $29\times30\times31$. The DOS is almost system-size independent. The IPR drops in the center of the impurity band.}
 \end{figure}

As analyzed in the previous section, with typical doping concentration $x=5\%$, we find that a clear impurity band in the DOS is developed at about $W=-4.5$. We plot the DOS and IPR together for different system sizes as shown in Fig.\ref{dosipr}.
The line of the DOS for different system sizes collapses to a single curve and it is the same as that from ED as shown in Fig.~\ref{DOSEvolution}, which tells us that we have already obtained the essential information of the impurity band.
As increasing the system size, the IPR does not change on the edge of the band which means the states on the edge of the whole band are localized. The IPR in the bulk decreases as enlarging the system,  especially at the center
of the impurity band ($E \sim -6.7$), the IPR drops to zero, which is the same as in the system bulk ($E \sim -4.0$). However, there is a small peak near $E\sim -5.5$ which is at the right edge of the impurity band. The IPR in the vicinity of this point tends to saturate to a fixed value as increasing the system size. The 
nonzero saturation of the IPR at this energy means another possible mobility edge existing near the junction between the conduction band and the impurity band.

\begin{figure}
 \includegraphics[width=8cm,height=10cm]{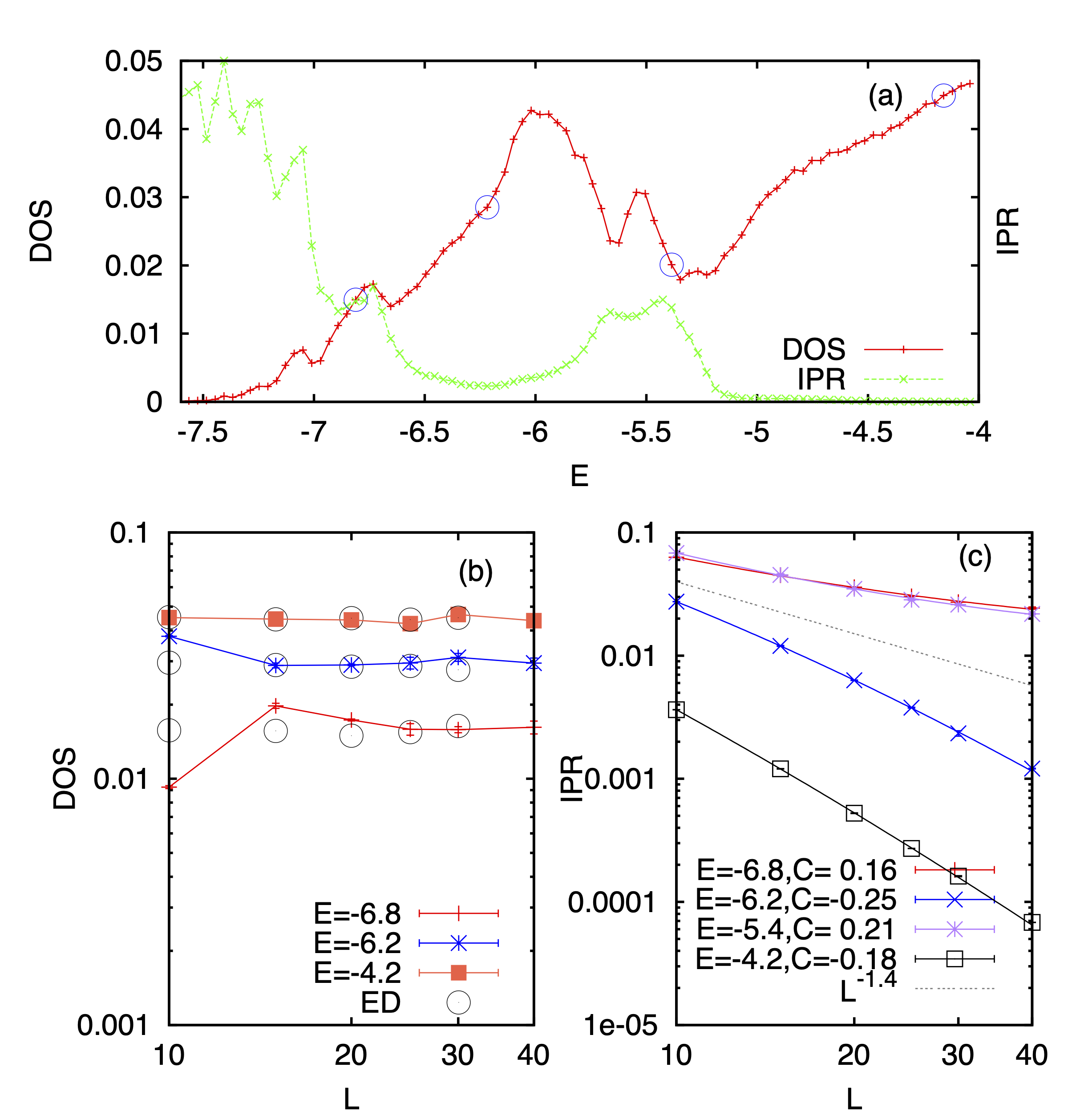}
\caption{\label{IPRL}The IPR/DOS as a function of system size for fixed energies. In fig.(c), we fit the data by using a function $\log(IPR) = A + B \log(L) + C \log(L)^2$. The curvature $C$ is labeled in the figure.}
\end{figure}

In order to justify our conjecture, we systematically study the value of IPR for several system sizes. As shown in Fig.~\ref{IPRL}(a), we choose four points from the knowledge of the DOS and IPR.
(1) $E=-4.2$ is in the bulk of the conduction band, at which the state is extended. (2) $E=-5.4$ is at the right edge of the impurity band. The state here is localized according to our conjecture. (3) $E=-6.2$ is in the bulk of the impurity band, which is extended according to its zero IPR value in large $L$ limit. (4) $E=-6.8$ is on the left edge of the impurity band and thus at the edge of the whole energy band. The state at the band edge is supposed to be localized. In Fig.~\ref{IPRL}(b) we again compare the DOS from JADA with that from Lapack which shows a convergence in large system size. According to the way of choosing these four points, (1) and (3) should have similar behavior as increasing the system size, and vice versa for (2) and (4).  Fig.~\ref{IPRL}(c) shows the IPR for these four energies in different system sizes. We plot the data in log scale and fit it by function
\begin{equation}
\log(\text{IPR}) = A + B \log(L) + C \log(L)^2.
\end{equation}
The sign of the curvature $C$ tells us whether the state is localized or not.
For (1) and (3), $C < 0$ means they are extended states, and oppositely $C > 0$ for localized states at points (2) and (4).

\begin{figure}
 \includegraphics[width=8cm]{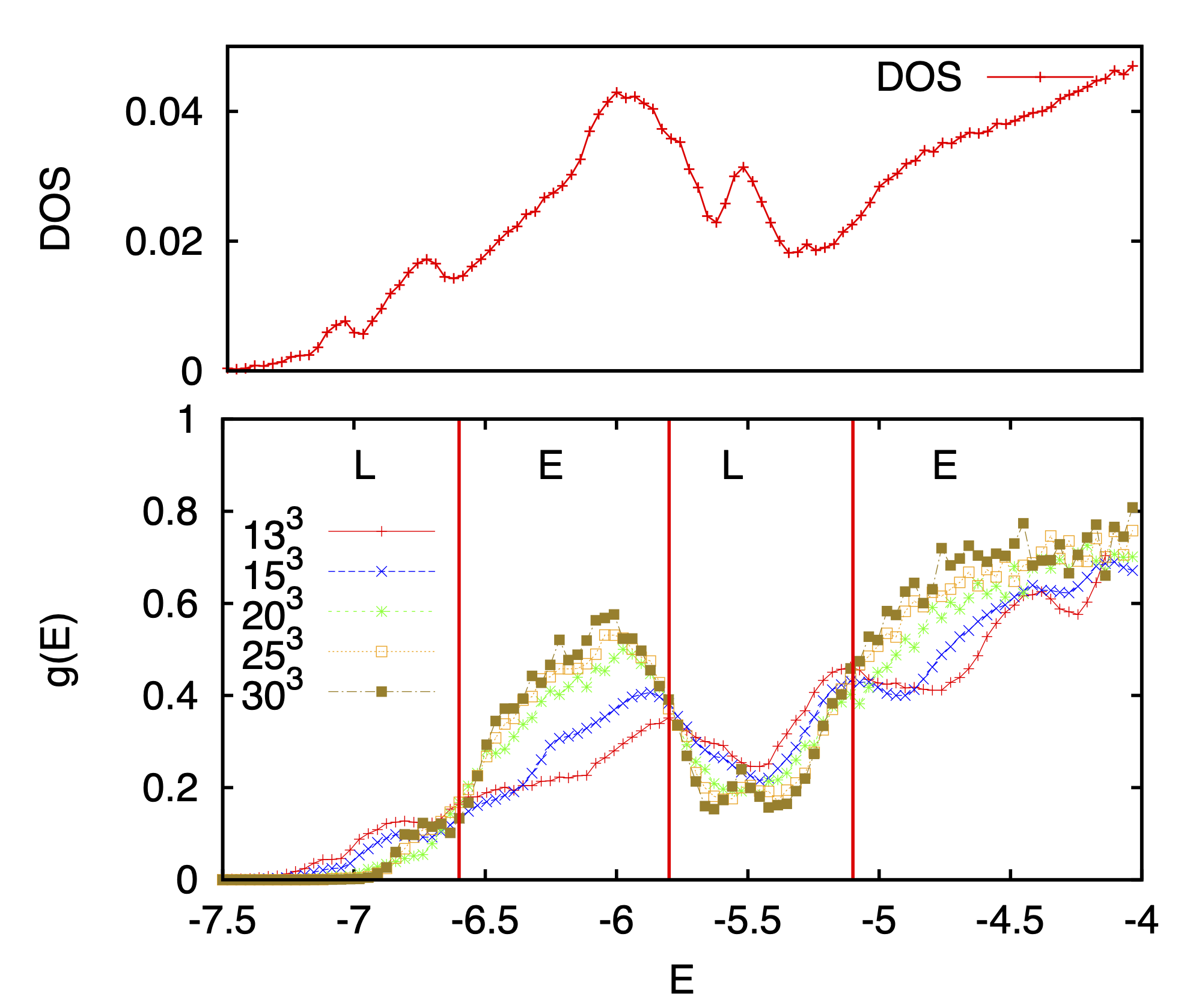}
\caption{\label{thouless}The Thouless number $g(E)$ as a function of energy for finite systems using exact diagonalization.}
\end{figure}

As another criterion, we calculate the Thouless number $g(E)$ for different system sizes. The results are shown in Fig.\ref{thouless} in which we plot the DOS together with the same horizontal axis. The impurity band has been divided into several regions at the crossover of $g(E)$ for different sizes. We label these regions by ``L" (localized) and ``E" (extended) to demonstrate different behavior $g(E)$.  As increasing the system size, it is obvious that the $g(E)$ increases in the ``E" region and decreases in the ``L" region. The energies with vertical lines are the locations of the mobility edges, or the boundaries between the localized states and extended states.  

\section{Deep learning approach} \label{sec:ml} 
Convolutional neural network(CNN), which is originally designed for 2D image recognition, has been widely adopted in studying phase transition and achieves high accuracy in recognition. A standard image recognition model can be used for a 3D electron system by integrating the 3D electron density in one direction. But the drawback of this approach is that the information of the electron density along one direction is lost during integration. So, we design a 3D CNN model for our 3D lattice model. To distinguish the localized and delocalized state, the CNN model will return two real numbers to represent the probability of the extended state $P$ and localized state ($1-P$) for the given wave function. If the probability of the extended state is larger than 0.5, we think the eigenstate is delocalized, and vice localized. Due to the limitation of the graphics memory (8GB) of our graphics card (NVIDIA GTX 1080), we consider a 3D $20\times20\times20$ lattice. The hidden layers in the CNN model consist of convolutional layers, max-pooling layers, and fully connected layers. The loss function is defined by the cross entropy $H(x) = -\sum_x p(x) \log q(x) $. During the training, we use the RMSPropOptimizer solver defined in Tensorflow\cite{tensorflow2015-whitepaper} as the stochastic gradient descent solver to minimize the loss function. The details of the neural network model are in Appendix \ref{app:arch}.

The training data for different phases are sampled from the 3-dimensional Anderson model using different disorder parameters. It's well known that the critical disorder at $E=0$ for the 3D Anderson model is 16.54 ±0.01~\cite{mackinnon1981one, mackinnon1983scaling, kramer1993localization}. When the disorder strength $W$ is larger than the critical value, the wave functions are exponentially localized and the system behaves as an insulator. Otherwise, the wave functions are delocalized and the system behaves as a metal. This phenomenon is known as Metal-Insulator Transition(MIT)\cite{anderson1958absence}. We get 4000 eigenstates from $W \in [14.0,16.0)$ as the delocalized phase and 4000 eigenstates from $ W \in [17.0,19.0)$  as the localized phase by steps of 0.1. For each W, we prepare 40 different realizations of randomness and for each realization, we take five eigenstates around $E = 0$. For the validation data set, we get another 600 eigenstates from $W \in [10.0,16.0) $ and 600 eigenstates from $W \in [17.0,23.0) $ in steps of 0.1. During each step of the training, we randomly select 256 eigenstates from the training data set as the input and calculate the gradient of the loss function with respect to the parameters in the CNN model and update them. After every 50 steps, we test the prediction accuracy on the validation data set and save the model with the highest prediction accuracy.
 
\begin{figure}[t]
    \centering
    \includegraphics[width=0.48\textwidth]{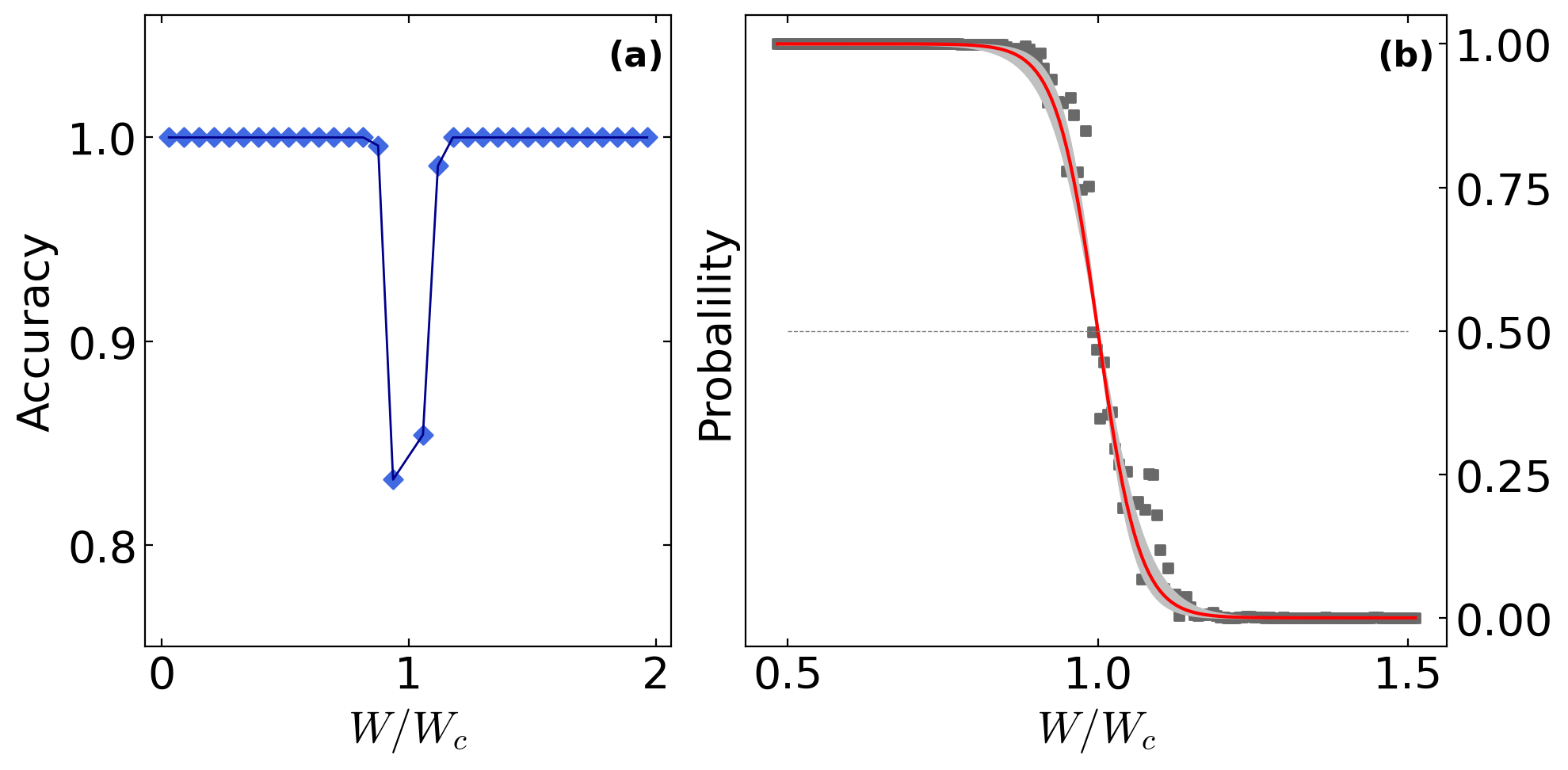}
    \caption{The performance of the trained neural network on Anderson model with different disorder parameters $W$. $W_c = 16.54$ is the critical disorder for $E=0$. (a) The classification accuracy of the trained neural network model. (b) The probability that the wave function is considered as an extended state by the trained neural network model. }
    \label{fig:accuracy}
\end{figure}

To show the prediction accuracy for different disorder parameters $W$, we generate another 16000 eigenstates sampled from the Anderson model using $W \in [0.1,16.0]$ and $W \in [17.0, 33.0)$. The prediction accuracy for different disorder strengths $W$ is shown in Fig.\ref{fig:accuracy}(a), and the overall accuracy is $99.0\%$. The lowest prediction accuracy around the critical disorder $0.8W_c < W < 1.2W_c$ is about $83\%$. We also test our trained model by producing the phase transition diagram of the 3D Anderson model. The testing data are sampled from $W \in [8.0,25.0]$ by steps of 0.1. In each realization of the same disorder parameter $W$, we pick 5 eigenstates around the band center($E=0$) as input data and use the averaged delocalized probability of the five eigenstates as the delocalized probability of this realization. We prepare 5 random realizations for each $W$ and average the delocalized probability. The phase diagram calculated using our trained CNN model is shown in Fig. \ref{fig:accuracy}(b). From Fig. \ref{fig:accuracy}(b), we see that the trained CNN model successfully captures the Metal-Insulator Transition(MIT). 

\begin{figure}[ht]
    \centering
    \includegraphics[width=8cm]{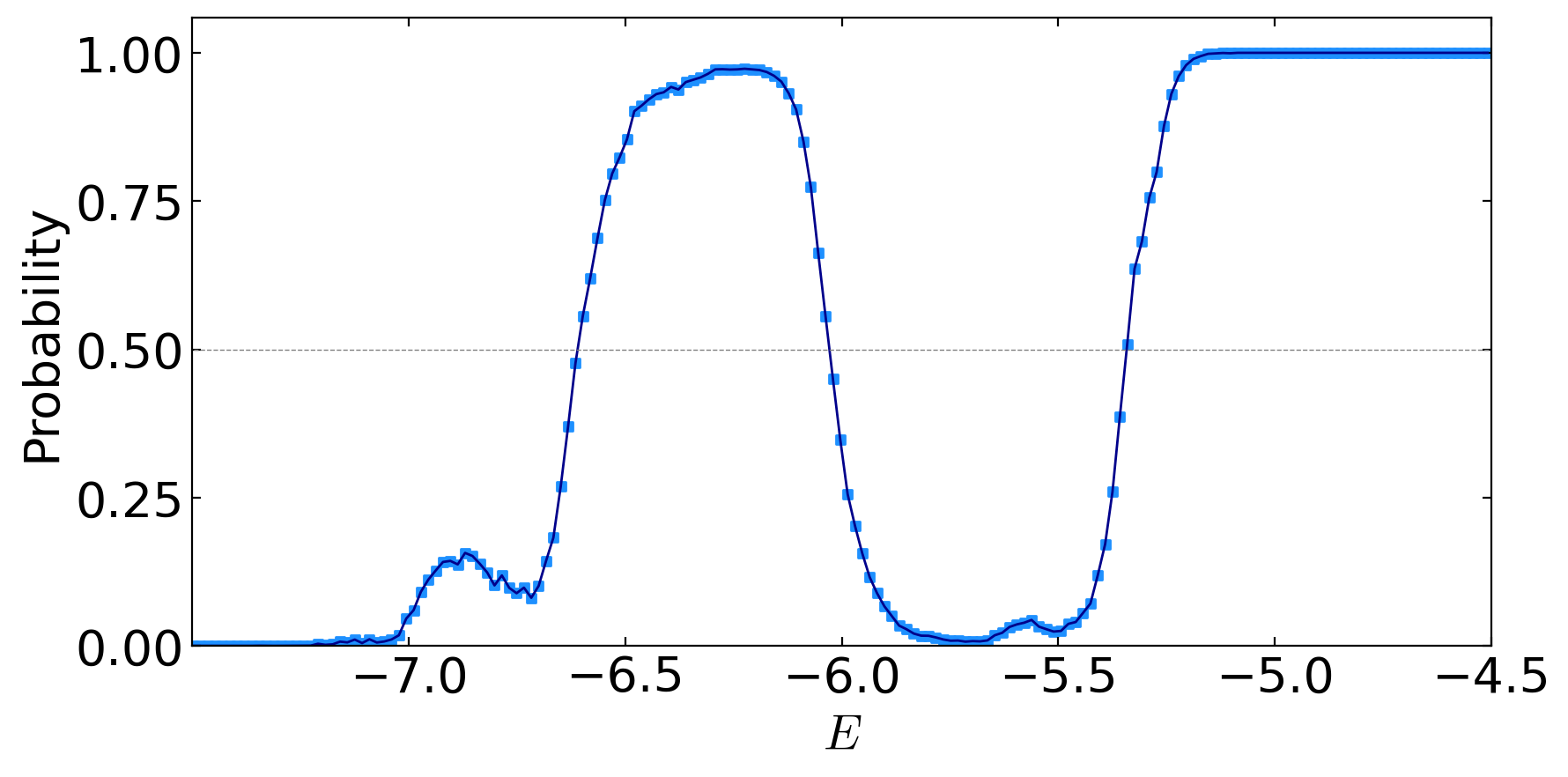}
    \caption{The probability that the corresponding wave function for different eigenenergies is considered as an extended state by the trained neural network model. The input wave functions are generated from Hamiltonian in Eq.\ref{eq:hamiltonian} using exact diagonalization. Averages over 1000 realizations are taken.}
    \label{fig:ml_impurity}
\end{figure}

Owing to its excellent classification accuracy, the trained neural network model is ready to find the extended state in the impurity band. We generate 1000 random realizations for the Hamiltonian in Eq.\ref{eq:hamiltonian} with doping probability $x=5\%$ and disorder parameter $W=-4.5$,  and obtain all eigenstates using the exact diagonalization method in Lapack. These quantum states are used as the input data for our trained CNN model to calculate the delocalized probability. We average the probability over 1000 realizations and the result is shown in Fig.\ref{fig:ml_impurity}. We can see that the CNN model confirms that delocalized states exist in the impurity band, which is in good agreement with the results obtained by IPR or Thouless number.

\section{Conclusions}
In this work, we numerically investigate the properties of the states in the ``impurity band'' of heavily-doped non-magnetic semiconductors. By using general full exact diagonalization and sparse matrix diagonalization with Jacobi-Division (JADA) method, we find that with a typical doping probability $x=5\%$, the impurity band in the DOS is developed at about $W=-4.5$. We calculate the IPR, Thouless number, and DOS together for different system sizes and study the relationship between them. The data fitting of IPR and system size on four points suggests the existence of the extended states in the impurity band. The Thouless number calculation supports the same conclusion and gives the exact location of mobility edges. 

Besides, we also utilize the supervised deep learning method, which is the state-of-the-art method in pattern recognition, to distinguish the extended and localized states in the impurity band. We train a 3D CNN model using the data generated from the Anderson model and then apply the trained neural network model to classify the states in the ``impurity band". Our trained neural network model achieves high accuracy ($99.0\%$) in classifying different states in the Anderson model. The prediction of our trained model on ``impurity band" also supports the finding from the relationship between IPR, Thouless number and system size though the predicted locations of mobility edges have small discrepancies. Our calculation gives direct evidence that there are three mobility edges in the impurity band for a specific on-site impurity potential in heavily-doped non-magnetic semiconductors.

\acknowledgements
Z-X. Hu is supported by the National Natural Science Foundation of China Grant No. 11974064 and 12147102, the Chongqing Research Program of Basic Research, and Frontier Technology Grant No. cstc2021jcyjmsxmX0081, Chongqing Talents: Exceptional Young Talents Project No. cstc2021ycjh-bgzxm0147, and the Fundamental Research Funds for the Central Universities Grant No. 2020CDJQY-Z003. HC acknowledges the U.S. Department of Energy, Office of Science, Basic Energy Sciences under Award No. DE-SC0022216.

\appendix
\section{Neural network model architecture and hyperparameters} \label{app:arch}

\begin{figure} 
    \includegraphics[width=0.3\textwidth]{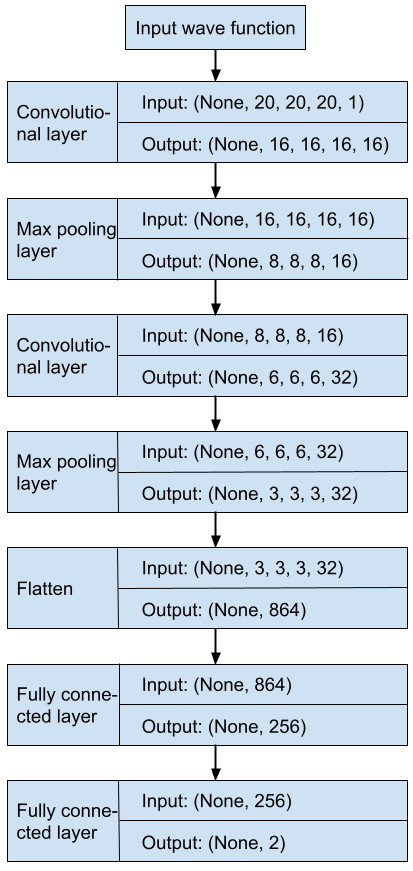}
    \caption{The architecture of the 3D CNN model employed in this paper for 3D $20\times20\times20$ lattice. ``\textit{None}" in the figure represents the batch size during training or evaluation, which is not a fixed number.  }
    \label{fig:cnn_arch}
\end{figure}

The 3D CNN model used in this paper has a similar architecture to the ``AlexNet"\cite{alexnet} and ``VGGNet"\cite{vggnet}, but with a smaller number of convolutional, max pooling, and fully connected layers. This is because we are dealing with a 3D lattice and the edges in the lattice have a much smaller length compared to the images. The architecture of our model is shown in Fig. \ref{fig:cnn_arch}, and the input and output dimension of each layer is also listed in the figure. 

The size of the convolution kernel applied in the first and second convolutional layers are $5\times5\times5$ and $3\times3\times3$, respectively. Activation function ReLU (rectified linear unit)\cite{nair2010rectified} is performed after the convolutional layer and fully connected layer except for the last layer, which is activated by the softmax\cite{bridle1989training} function. Bias parameters are included for artificial neurons. Dropout\cite{srivastava2014dropout} is performed with probability $p=0.5$ after the first fully connected layer to avoid over-fitting and increase the evaluation accuracy.

\bibliography{reference}
 
\end{document}